\documentclass[prd,twocolumn,a4paper,superscriptaddress,floatfix]{revtex4}
\usepackage{graphicx}
\begin{document}
\newcommand{\be}{\begin{equation}}
\newcommand{\ee}{\end{equation}}
\newcommand{\bq}{\begin{eqnarray}}
\newcommand{\eq}{\end{eqnarray}}
\newcommand{\bsq}{\begin{subequations}}
\newcommand{\esq}{\end{subequations}}
\newcommand{\bc}{\begin{center}}
\newcommand{\ec}{\end{center}}
\newcommand\lapp{\mathrel{\rlap{\lower4pt\hbox{\hskip1pt$\sim$}} \raise1pt\hbox{$<$}}}
\newcommand\gapp{\mathrel{\rlap{\lower4pt\hbox{\hskip1pt$\sim$}} \raise1pt\hbox{$>$}}}
\def\E{{\cal E}}

\title{Evolution of Cosmic Necklaces and Lattices}
\author{C.J.A.P. Martins}
\email[Electronic address: ]{Carlos.Martins@astro.up.pt}
\affiliation{Centro de Astrof\'{\i}sica, Universidade do Porto, Rua das Estrelas, 4150-762 Porto, Portugal}
\affiliation{CTC, DAMTP, University of Cambridge, Wilberforce Road, Cambridge CB3 0WA, United Kingdom}
\date{28 July 2010}
\begin{abstract}
Previously developed analytic models for the evolution of cosmic string and monopole networks are applied to networks of monopoles attached to two or more strings; the former case is usually known as cosmic necklaces. These networks are a common consequence of models with extra dimensions such as brane inflation. Our quantitative analysis agrees with (and extends) previous simpler estimates, but we will also highlight some differences. A linear scaling solution is usually the attractor solution for both the radiation and matter-dominated epochs, but other scaling laws can also exist, depending on the universe's expansion rate and the network's energy loss mechanisms.
\end{abstract}
\pacs{98.80.Cq, 11.27.+d, 98.80.Es}
\maketitle

\section{\label{intro}Introduction}

Currently favored fundamental theories suggest that we live on a four-dimensional slice of a higher-dimensional universe, and while most forces are confined to our slice, gravity may leak off. These so-called brane world models may provide natural explanations for inflation (here known as brane inflation), and topological defects may at the end of it. The type of defect network which is formed and its basic properties (from its energy scale to whether or not it is long-lived) will depend on the specific details of the model in question \cite{KIBBLE}. A thorough overview of the subject may be found in the book by Vilenkin and Shellard \cite{VSH} and in more recent review articles \cite{REV1,REV2,REV3}.

Most of the past work on defects concerns the simplest models of cosmic strings, on the grounds that they are cosmologically benign, and are a generic prediction of inflationary models based on Grand Unified Theories \cite{Rachel1,Rachel2} or branes \cite{Branes1,Branes2}, while domain walls and monopoles tend to be cosmologically dangerous and tightly constrained. However, it is clear, particularly in the context of models with extra dimensions, that networks containing more than one type of defect will often be produced. Two examples that have attracted considerable interest are semilocal strings \cite{SEMILOCAL1,SEMILOCAL2,SEMILOCALSIM} and cosmic necklaces \cite{HINDMARSH}. For the latter these claims have been made both in the context of brane inflation \cite{Dasgupta1,Chen,Dasgupta2} and in string theory itself \cite{LEBLOND,LAKE}.

This is the third report on an ongoing project which is addressing some of these issues. In the past we have developed \cite{MONOPOLES} an analytic model for the evolution of networks of local and global monopoles \cite{DIRAC,THOOFT,POLYAK}. The model is analogous to the velocity-dependent one-scale model for cosmic strings \cite{MS0,MS1,MS2}, which has been extensively tested against field theory \cite{ABELIAN,MS3} and Goto-Nambu simulations \cite{MS3,MS4}. This was then extended \cite{HYBRIDS} to the case of monopoles attached to one string (the so-called hybrid networks \cite{VILENKIN}), as well to vortons \cite{VORTONS,MSVORTONS}. Here we study defect networks where monopoles are attached to two or more strings. Networks of the first type are commonly called cosmic necklaces; we shall refer to networks of of monopoles attached to three or more strings as cosmic lattices.

The behavior of necklaces and lattices is qualitatively similar, and will be for the most part treated together in this paper, through we will point out the small existing differences. However, their evolution differs in several key aspects from both that of individual monopoles and that of monopoles attached to a single string (usually called hybrid networks) \cite{MONOPOLES,HYBRIDS}. The main difference is that necklaces and lattices form stable, long-lived networks which usually reach a scaling solution.

\section{\label{review}Cosmic Necklaces and Lattices}

We start with a brief overview of previous results on the evolution of necklaces and lattices. This is by no means exhaustive; the aim is to highlight the dynamical aspects we will need to model. A more detailed discussion can be found in \cite{VSH} as well as in other earlier references that we will point out where appropriate.

The defect networks of interest form via the symmetry breaking pattern $G\to K\times U(1)\to K\times Z_N$. If $G$ is  a semi-simple group, the first phase transition produces monopoles while in the second each monopole becomes attached to $N$ strings. If $K$ is trivial all the (Abelian) magnetic flux of the monopoles is confined into the strings, and there are no unconfined fluxes. However, unconfined non-Abelian magnetic fluxes can exist in the generic case. The previously discussed hybrid case \cite{HYBRIDS} corresponds to $N=1$. Here we will discuss the case $N\ge2$; $N=2$ corresponds to cosmic necklaces and $N\ge3$ to cosmic lattices.

The corresponding defect masses will be $m\sim ({4\pi}/{e})\eta_m$ and $\mu\sim 2\pi\eta^2_s$, while the characteristic monopole radius and string thickness are $\delta_m\sim (e\eta_m)^{-1}$ and $\delta_s\sim (e\eta_s)^{-1}$. There are also scenarios where the intermediate phase transition is absent, $G\to K\times Z_N$, in which case an analogous network still forms but the role of the monopoles is now played by solitons that are usually called 'beads'. In this case the two energy scales are obviously similar, that is $\eta_s\sim\eta_m$.

Up to the second transition (if it exists) the models for plain monopoles \cite{MONOPOLES} apply, but once the strings form a separate treatment is needed. Two key differences are immediately apparent. The evolution of isolated monopoles can be divided into a 'free' (pre-capture) and a post-capture period, with captured monopoles effectively decoupling from the network and losing energy radiatively until they decay (this is analogous to the evolution of cosmic string loops). In the present context the monopoles are effectively captured by the strings, and one needs to account explicitly not only for radiative losses (for example gauge radiation if there are unconfined magnetic charges) but also for the force the strings exert on the monopoles---depending on the context, the forces due to the string(s) or the other monopoles may be the dominant ones.

If all the strings attached to each monopole have the same tension (which we will assume to be the case in the present paper) then all the strings pull it with equal forces, and therefore there is no tendency for a monopole to be captured by the nearest antimonopole, unless their separation is of order $\delta_s$. If there are $N$ strings attached to each monopole, its proper acceleration is given by the vector sum of the tension forces exerted by the strings. At a back-of-the-envelope level, each force is of order $f\sim\mu$, and hence one expects that $a\sim\mu/m$. Monopoles should therefore be accelerated to relativistic speeds provided that the characteristic length of string segments, $L_s$, is such that $\mu L_s\gg m$, that is ${L_s}/{\delta_s}\gg{\eta_m}/{\eta_s}$.

Aryal et al. \cite{ARYAL} first studied the formation and statistical properties of these networks, for $N=2$ and $N=3$, showing that for $N\ge 3$ a single network is formed. In all cases they find that the system is dominated by one infinite network comprising more than $90\%$ of the string length. Some finite networks and closed loops do exist, in numbers rapidly decreasing with their size. Finally, most of the string segments have a length comparable to the typical distance between monopoles (much larger segments being exponentially suppressed). This justifies our assumption of an inter-monopole separation, $L_m$, comparable to $L_s$.

The cosmological evolution of these networks was first discussed by Vachaspati \& Vilenkin \cite{VACHVIL}, who argued that assuming that the radiation of gauge quanta is the dominant energy loss mechanism of the networks, they reach scaling with a characteristic lengthscale $L\sim ({\eta^2_s}/{\eta^2_m}) t$ and the monopoles become highly relativistic. These networks can also lose energy by producing closed loops of string and small nets. The effect of these is harder to estimate, but as we shall discuss it is fairly easy to model phenomenologically..

Specifically they divide the network's energy into string and monopole parts, with $\rho_m=\beta\rho$, $\rho_s=(1-\beta)\rho$ and an effective equation of state $3{p}/{\rho}=\beta+(1-\beta)(2v^2_s-1)$ with $V_s$ being the string velocity and assuming $v_m\sim1$. The evolution equation is then
\be\label{vvdynamics}
{\dot\rho}=-3H(\rho+p)-nw
\ee
where $w\sim{(ga)^2}/{6\pi}$ for gauge radiation losses and $L\sim n^{-1/3}$ is both the length of string segments and the average monopole distance. Assuming a self-similar evolution one can also set $\rho_s\sim\mu n^{2/3}$.

For the case without unconfined magnetic fluxes (that is, no radiation), they claim $L\propto t^\alpha$, expecting $\alpha<1$: there is no scaling and the defects eventually dominate the energy density of the universe. But note that in saying this they are specifically thinking of the radiation epoch (it is clear that the behavior of their solution depends on several parameters including the expansion rate).

The specific case of cosmic necklaces has subsequently been studied by Brezinsky \& Vilenkin \cite{BEREVIL}. They assume no unconfined magnetic fluxes (hence no Coulomb forces between the monopoles) and characterize the networks by a dimensionless ratio $r=m/(\mu L)$ with the average mass per unit length of the necklaces being $(r+1)\mu$. They also neglect the effect of annihilations (though the validity of this assumption has been challenged \cite{SIEMENS,BLANCOPILLADO}), and find that the system tends to evolve towards large $r$. Again the necklaces are expected to evolve in a scaling regime, with a characteristic network lengthscale $\xi$. The force per unit length of string is $f\sim{\mu}/{\xi}$ and the acceleration is $a^{-1}\sim (r+1)\xi$ so we expect that
\be\label{berez3}
\xi\sim\frac{t}{\sqrt{1+r}}\,,\quad v\sim\frac{1}{\sqrt{1+r}}\,.
\ee
In the limit $r<<1$ the monopoles are sub-dominant and the strings will behave approximately as ordinary ones. In the limit $r>>1$ the strings are very slow and their separation is small. This is a very simple toy model, as in fact $r$ is generically not a constant parameter, since $d\sim\xi$, so these solutions are only approximate. Nevertheless, this approach has the advantage of algebraic simplicity, and we shall show below that in appropriate circumnstances it can be related to more robust models. Analogous results have been found with a somewhat different toy model \cite{LEBLOND}.

\section{\label{models}Quantitative Evolution}

It's easy to start modeling these networks by using the evolution equation derived in our previous work \cite{MONOPOLES,HYBRIDS}. Most authors at this point focus on the evolution of the strings, treating the monopoles (as it were) as a small correction. Our approach, justified in \cite{HYBRIDS}, is precisely the opposite---we focus on the evolution of the separation between monopoles.

For this context the evolution equations for the characteristic separation $L$ and root-mean squared velocity $v$ of the monopoles are
\be\label{modell}
3\frac{dL}{dt}=(3+v^2)HL+Q_\star
\ee
\be\label{modelv}
\frac{dv}{dt}=(1-v^2)\left(k_s\frac{\eta^2_s}{\eta_m}-Hv\right)\,.
\ee

We have neglected the term due to friction in both equations (it's easy to show this is subdominant at late times). The energy loss term $Q_\star$ in the lengthscale equation is a renormalized quantity, accounting for the various losses present (including string intercommutings, radiation and annihilations), as discussed in our previous work. There may be a velocity-dependence of some of these contributions, but as we shall see monopoles will typically have ultra-relativistic velocities $v\sim1$ and therefore this dependence can be neglected. The velocity equation includes the force due to the strings (with a phenomenological curvature parameter $k_s$ that is discussed in \cite{MS2}) but we have neglected that due to monopoles, since if it exists (which is only the case for unconfined fluxes) it's always smaller than that due to the strings. Indeed, using the above definitions of mass scales and thicknesses one finds that the ratio of the two forces is
\be
\frac{f_m}{f_s}\sim\frac{k_m}{k_s}\left(\frac{\delta_s}{L}\right)^2\ll1\,;
\ee
we expect the $k_i$ to be (dimensionless) coefficients of order unity, though note that they should be different for necklaces and lattices.

From the velocity equation we immediately confirm that the monopole velocities will be driven towards unity, $v\to1$, as previously stated. As for the monopole lenght scale, assuming a generic expansion rate $a\propto t^\lambda$, we find two different regimes for slow and fast expansion rates
\be
L=\frac{Q_\star}{3-4\lambda}t\,,\quad \lambda<3/4
\ee
\be
L\propto a^{4/3}\propto t^{4\lambda/3}\,,\quad \lambda\ge 3/4\,.
\ee
The former explicitly requires a non-zero energy loss rate---we will return to this point later. We therefore have linear scaling both in the radiation and matter eras (as generically claimed by previous authors, based on simpler qualitative arguments). For monopoles $L\propto t$ corresponds to the monopole density decreasing relative to that of the background. However, for fast expansion rates the growth is superluminal, and the network will eventually disappear. An analogous scaling solution was already discussed in our previous work on hybrid networks.

Its easy to establish a link between this analysis and that of Vachaspati \& Vilenkin \cite{VACHVIL} (which is embodied in Eq. \ref{vvdynamics}) and thus to carry out a more detailed analysis of of the possible scaling solutions. Let's apply Eq. \ref{vvdynamics} to the monopoles. We use both the notation and definitions of our previous work and those of \cite{VACHVIL}. Since $\rho_m=mn={m}/{L^3}$ and the monopole equation of state is $3p=v^2_m\rho$, we get by substitution
\be\label{modelmon}
3\frac{dL}{dt}=(3+v^2)HL+\frac{L}{\eta_m}w\,.
\ee
Now, Vachaspati \& Vilenkin are assuming energy losses through gauge radiation; noting that
\be
w\sim\frac{(ga)^2}{6\pi}\sim\left(\frac{\mu}{\eta_m}\right)^2\sim{\dot\epsilon}_{gauge}
\ee
\be
Q_{gauge}\sim L\frac{{\dot\epsilon}_{gauge}}{{\epsilon}_{gauge}}\sim\frac{L}{\eta_m}w
\ee
we see that this evolution equation for $L$ is exactly the same as Eq. \ref{modell}, matching the $Q$ terms (which in our case can phenomenologically account for further energy loss channels).

We can also apply Eq. \ref{vvdynamics} to the strings. In this case $\rho_s=\mu n^{2/3}={\mu}/{L^2}$ and the string equation of state is $3p=(2v^2_s-1)\rho$; again we find
\be\label{modelstr}
2\frac{dL}{dt}=2HL(1+v^2_s)+{w}{\mu}\,.
\ee
In this case we have
\be
{w}{\mu}\sim \left(\frac{\eta_s}{\eta_m}\right)^2\sim Q\,.
\ee
Note the interesting fact that the dimensionless parameter $Q$ determines the energy loss term for both the strings and the monopoles. The above is the usual evolution equation for the cosmic string correlation length \cite{MS0,MS1,MS2}, if one assumes a constant string velocity---otherwise the energy loss term should depend linearly on velocity.

The scaling solution for the monopoles has already been discussed. For the case of the strings the solution is also the expected linear scaling
\be
L=\frac{Q}{2-2\lambda(1+v^2_s)}t\,,
\ee
for constant velocities and provided $\lambda(1+v^2_s)<1$. But this solution is an attractor, as in the case of normal strings: for very large lengthscales the string velocity would no longer be a constant (the string velocity evolution equation would drive it to smaller values), and a new equilibrium value with a smaller lengthscale would be reached.

We can also consider more generic scaling solutions of the above equations, ie allowing for the possibility of zero energy losses ($Q=0$). We will confirm that scaling ($L\propto t$) generically requires $Q\neq0$. Starting again with the monopoles, for $Q\neq0$ we have the two branches of the solution discussed above. For $Q=0$ the solution is always
\be
L\propto a^{4/3}\propto t^{4\lambda/3};
\ee
note that for $\lambda<3/4$ the lengthscale grows subluminally while for $\lambda>3/4$ it grows superluminally. In the absence of radiative energy losses, only a fast enough expansion can dilute the network. We can also compare the evolution of the monopole and background densities
\be
\frac{\rho_m}{\rho_b}\sim\frac{\eta_m}{m^2_{Pl}}\frac{t^2}{L^3}\,.
\ee
For the linear scaling solution $L\propto t$ this has the form 
\be
\frac{\rho_m}{\rho_b}\sim\frac{1}{Q^3_\star}\left(\frac{\eta_m}{m_{Pl}}\right)^2\left(\frac{T}{m_{Pl}}\right)^2\propto\frac{1}{t}\,.
\ee
From this we see that if gauge radiation is present then the energy density of the network is smaller than the background density. However, if the only radiative channel available is gravitational radiation (in which case, as discussed in \cite{HYBRIDS}, $Q_\star=Q_{grav}\sim(\eta_s/m_{Pl})^2$) then the network energy density is in fact the dominant one. For the non-scaling branch the density is
\be
\frac{\rho_m}{\rho_b}\propto t^{2-4\lambda}\,,
\ee
and the behavior depends on the cosmological epoch. Notice that during the radiation era the density is a constant fraction of that of the background.

These results can now be related with the toy-model analysis of Berezinsky \& Vilenkin. In the linear scaling regime we have $r=\rho_m/\rho_s\propto t^{-1}$ and therefore $r\to0$ and $v\to1$ in agreement with our quantitative analysis. In the non-scaling branch $r\propto t^{2-4\lambda}$, whose behavior again depends on the cosmological epoch. The radiation epoch corresponds to the interesting case $r=const.$, while faster expansion rates (say, the matter-dominated epoch) dilute the monopole density relative to that of strings. Slower expansion rates ($\lambda<1/2$), which sometimes occur in the very early universe in string cosmology models, would have a growing density ratio $r$.

\section{\label{concl}Summary and outlook}

We have extended a recently developed analytic model for the evolution of monopole \cite{MONOPOLES} and hybrid networks \cite{HYBRIDS} to the case of monopoles attached to several strings. We discussed their possible scaling solutions, generically confirming the expectation that the network will reach linear scaling (with its characteristic lengthscale $L\propto t$), but also showing that other scaling behaviors can occur, depending on the expansion rate of the universe and on the energy loss mechanisms available to the network.

This completes the basic structure of analytic tools needed to study defect networks containing both strings and monopoles. These models have the advantage of conceptual simplicity, in addition to that of allowing a quantitative description of the network's evolution. Nevertheless, a further generalization will be required in order to more accurately describe the possible effects of a hierarchy of string tensions \cite{TYE,LEBLOND,TASOS}. A further interesting case, which requires additional dynamics  is that of semilocal strings \cite{SEMILOCAL1,SEMILOCAL2}, which we will address in a subsequent publication.

\begin{acknowledgments}
The work of C.M. is funded by a Ci\^encia2007 Research Contract, funded by FCT/MCTES (Portugal) and POPH/FSE (EC). I am grateful to Ana Ach\'ucarro for valuable discussions and collaboration on related issues, and to CTC, Cambridge (where part of this work was done) for the hospitality.
\end{acknowledgments}

\end{document}